\documentclass[preprint,amsmath,amssymb,aps,floatfix,pra,letter]{revtex4-1}

\usepackage{graphicx}
\usepackage{dcolumn}
\usepackage{bm}
\usepackage{float}
\usepackage{color}

\begin{document}


\title{Linear and quadratic static response functions and \\ structure functions in Yukawa liquids}

\author{P. Magyar$^1$, Z. Donk\'o$^{1,2}$, G. J. Kalman$^2$, K. I. Golden$^3$}
\affiliation{$^1$Institute for Solid State Physics and Optics,
Wigner Research Centre for Physics,\\
Hungarian Academy of Sciences, H-1121 Budapest, Konkoly-Thege Mikl\'os str. 29-33, Hungary}
\affiliation{$^2$Physics Department, Boston College, Chestnut Hill, MA 20467, USA}
\affiliation{$^3$Department of Mathematics and Statistics, and Department of Physics, University of Vermont, Burlington, USA}

\date{\today}

\begin{abstract}
We compute linear and quadratic static density response functions of three-dimensional Yukawa liquids by applying an external perturbation potential in molecular dynamics simulations. The response functions are also obtained from the equilibrium fluctuations (static structure factors) in the system via the fluctuation-dissipation theorems. The good agreement of the quadratic response functions, obtained in the two different ways confirms the quadratic fluctuation-dissipation theorem. We also find that the three-point  structure function may be factorizable into  two-point structure functions, leading to a cluster representation of the equilibrium triplet correlation function. 
\end{abstract}


\maketitle

\section{Introduction}

The response of many-particle systems to external perturbations has been a topic of continuous interest. The linear fluctuation-dissipation theorem (FDT) establishes a relationship between equilibrium two-point correlations in a system and its linear response to a small external perturbation. With increasing amplitude of perturbation, however, the response of the system may become nonlinear, in which regime the second or higher order (nonlinear) response functions play a role. An extension of the conventional FDT to this regime, the {\it quadratic fluctuation-dissipation theorem} (QFDT), formulated first some time ago by Golden, Kalman and Silevitch \cite{paper1}  and Sitenko \cite{Sitenko}, establishes a relationship between the quadratic response functions and the equilibrium three-point correlations. The logical extension of the FDT providing the fundamental link between the $n$-th order response functions (non-equilibrium transport coefficients) and their companion $(n+1)$-point equilibrium correlations of fluctuating quantities leads to the notion of the hierarchy of nonlinear fluctuation-dissipation theorems.  This has become a topic studied by a number of investigators representing a wide range of disciplines, most notably, condensed matter and plasma physics \cite{paper1,cmp}, nonlinear optics \cite{nop}, high energy physics \cite{hep}, chemistry \cite{chem}, statistical physics \cite{statphys}, and many-body physics \cite{mbp}. However, while the linear FDT is well established and thoroughly tested, the testing of the quadratic FDT, either experimentally or via simulations, has been missing so far.  Addressing this issue is the motivation for this work.

A concrete system, which we will focus on is the one-component plasma-like model, in which particles interact via a screened Coulomb (Debye-H\"uckel, or Yukawa) potential (Yukawa system). Physical systems lending themselves to the approximation of the interaction by such a potential are charged colloids \cite{colloids} and dusty (complex) plasmas \cite{dusty}. In SI units, the inter-particle potential reads
\begin{equation}
\phi(r) = \frac{Q}{4 \pi \varepsilon_0} \frac{\exp(-r/\lambda_D)}{r},
\end{equation}
where $Q$ is the charge of the particles and $\lambda_D$ is the screening (Debye) length. The ratio of the inter-particle potential energy to the thermal energy is expressed by the coupling parameter
\begin{equation}
\Gamma = \frac{Q^2}{4 \pi \varepsilon_0 a k_B T},
\end{equation}
where $T$ is temperature. We introduce the screening parameter $\kappa = a / \lambda_D$, where $a = (3/4 \pi n)^{1/3}$ is the Wigner-Seitz radius and $n$ is the particle number density. We investigate the system in the {\it strongly coupled liquid phase} ($\Gamma \gg 1$), where a prominent liquid structure builds up. It is not the subject of our studies, but we note that the system turns into a crystal when the coupling parameter reaches a certain value that depends on the screening parameter \cite{Hamaguchi}.
In the following we investigate the static linear and quadratic longitudinal responses of the Yukawa system and the related equilibrium two-point and three-point correlations.

We adopt the molecular dynamics (MD) simulation approach for our investigations. The effect of an external potential can be implemented in the simulation in a straightforward manner: in addition to the inter-particle forces, each particle is exposed to an external force ${\bf F}_{ext}({\bf r}) = - \nabla \hat{\Phi}({\bf r})$. (Variables marked with ``hat'' correspond to {\it external} quantities, which are distinguished from {\it total} quantities, e.g. the total electric field is composed of an external field plus the polarization field.) The external potential energy $\hat{\Phi}({\bf r})$ results in the development of a perturbed density profile of the plasma:
\begin{equation}
n({\bf r}) = n_0 + \tilde{n}({\bf r}),
\end{equation}
which is linked to the external perturbation via the density response function. In general, the deviation $\tilde{n}({\bf r}$) from the homogeneous density $n_0$, in the Fourier (wave number) space is given by the following perturbation series:
\begin{eqnarray}
\langle \tilde{n}(\textbf{k}_0) \rangle = \sum_{s=1}^{\infty}\bigg\{ \frac{1}{V^{s-1}} \sum_{\textbf{p}_{1}...\textbf{p}_{s}}\hat{\chi}^{(s)}(\textbf{p}_{1},...,\textbf{p}_{s}) \times
\nonumber \\
\hat{\Phi}(\textbf{p}_{1})...\hat{\Phi}(\textbf{p}_{s}) ~\delta_{\textbf{p}_{1}+...+\textbf{p}_{s}+\textbf{k}_0,0} \bigg\}
\label{sor}
\end{eqnarray}
where $V$ is the volume of the system, $\hat{\chi}^{(s)}(\textbf{p}_{1},...,\textbf{p}_{s})$ is the $s$-th order {\it external response function}, and $\langle ~ \rangle$ denotes ensemble average.

Measuring the perturbed density profile $\langle n(\textbf{r}) \rangle$ induced by a given external potential energy $\hat{\Phi}$ allows us to determine the response function $\hat{\chi}$. Our goal is to determine the linear and quadratic response functions, $\hat{\chi}^{(1)}({\bf k}_1)$ and $\hat{\chi}^{(2)}(\textbf{k}_{1},\textbf{k}_{2})$. Finding the linear $\hat{\chi}^{(1)}({\bf k}_{1})$ -- a function of a single variable --  can be done with relative ease. In the quadratic case, however, we deal with a large parameter space of three scalar variables $k_1$ and $k_2$, and the angle between $\textbf{k}_{1}$ and $\textbf{k}_{2}$. Rather than attempting to address the immense task of mapping the entire parameter space we restrict the simulation to analyzing  a few representative one-parameter samples:  we apply (i) a single harmonic perturbation and (ii) a superposition of two harmonic perturbations (a ``biharmonic perturbation'') such that  $\textbf{k}_{1}$ and $\textbf{k}_{2}$ are related to each other by a few chosen constraints. We trust that sufficient information is gained in this way to reveal the salient features of the problem. 

\section{Theoretical background}

In this section we will establish the connection between the external perturbation potential (energy) and the induced density response. In the case of the harmonic perturbation the derivation is given in details, for the biharmonic case these details are omitted.

\subsection{Fluctuation-dissipation theorems}

The linear FDT connects the linear response function with the static structure function through the relationship

\begin{equation}
\hat{\chi}^{(1)}(\textbf{k}_1)=-\beta n_{0} S(\textbf{k}_1),
\label{eq:fdt1}
\end{equation}
while the quadratic FDT establishes a relationship between the quadratic response function and the second order structure function:
\begin{equation}
\hat{\chi}^{(2)}(\textbf{k}_{1},\textbf{k}_{2};\textbf{k}_{0}) = \frac{\beta^{2}n_{0}}{2} S(\textbf{k}_{1},\textbf{k}_{2};\textbf{k}_{0}),
\label{eq:fdt2}
\end{equation}
where 
\begin{equation}
\textbf{k}_{1}+\textbf{k}_{2}+\textbf{k}_{0}=0.
\end{equation}
Here $\beta = 1 / kT$. 

The $S(\textbf{k}_1)$ and $S(\textbf{k}_{1},\textbf{k}_{2})$ static structure functions are defined as:
\begin{equation}
S(\textbf{k}_1)=\frac{1}{N}\langle n(\textbf{k}_1,t) n(-\textbf{k}_1,t)\rangle_{(t)}
\label{eq:linsk}
\end{equation}
and
\begin{equation}
S(\textbf{k}_{1},\textbf{k}_{2};\textbf{k}_{0})=\frac{1}{N}\langle n(\textbf{k}_{1},t) n(\textbf{k}_{2},t) n (\textbf{k}_{0},t)\rangle_{(t)},
\label{eq:quadsk}
\end{equation}
where $N$ is the number of particles, and $n(\textbf{k},t)$ is the microscopic density in Fourier space:
\begin{equation}
n(\textbf{k},t)=\sum_{j=1}^{N}e^{-i\textbf{k}\textbf{r}_{j}(t)}.
\end{equation}

A corollary to the FDT-s is a relationship between the static structure functions and the correlation functions of the system. We start with the conventionally defined two-particle and three-particle distribution functions  $g({\bf r}_1,{\bf r}_2)$ and $g({\bf r}_1,{\bf r}_2,{\bf r}_3)$, and define the respective two-particle and three-particle correlation functions by
\begin{equation}
g({\bf r}_1,{\bf r}_2) \equiv g({\bf r}_{12}) = 1+h({\bf r}_{12})
\end{equation}
and 
\begin{eqnarray}
& g({\bf r}_1,{\bf r}_2,{\bf r}_3) \equiv g({\bf r}_{12}, {\bf r}_{23}) = \nonumber \\
& 1+h({\bf r}_{12})+h({\bf r}_{23})+h({\bf r}_{31})+h({\bf r}_{12}, {\bf r}_{23}).
\end{eqnarray}	
Their Fourier transforms are linked  to the respective structure functions by the well-known linear
\begin{equation}
S({\bf k}_1) = 1+ n h({\bf k}_1)
\label{s1}
\end{equation}
and quadratic \cite{paper1}
\begin{eqnarray}
& S({\bf k}_1,{\bf k}_2;{\bf k}_0) =   \nonumber \\
& 1+ n [ h({\bf k}_1)+ h({\bf k}_2)+ h({\bf k}_0)] + n^2 h({\bf k}_1,{\bf k}_2)
\label{s2}
\end{eqnarray}
relationships.

In the above definitions, the structure functions are given as ensemble averages at an arbitrary time, we indicate this by the notation $\langle \rangle_{(t)}$. In the simulations these functions are calculated via time averaging of the data obtained in subsequent time steps for a finite system of $N$ particles. Using time averaging rather than ensemble averaging is justified by the ergodicity of the system.

We note that, strictly speaking, the above FDTs (\ref{eq:fdt1}), (\ref{eq:fdt2}) and definitions (\ref{sor}), (\ref{eq:linsk}), (\ref{eq:quadsk}), (\ref{s1}), (\ref{s2}) are valid only as long as all the wavenumber variables differ from zero. In the $k_i=0$ domains the linear and quadratic response functions and the two- and three-point structure functions exhibit a singular behavior. In the following, we are interested in perturbations and responses at finite $k$ values only: consequently the question of what happens at $k_i=0$ will be ignored. 

\subsection{Harmonic perturbation}

Here we will derive the response of the system to a form of external potential:
\begin{equation}
\hat{\Phi}(\textbf{r}) = c f_{0} \cos(\textbf{k}_{1}\textbf{r}).
\label{pot1}
\end{equation}
In this form $f_0$ is the degree of perturbation, while $c$ is an additional parameter that ensures that the maximum of the external force at $f_0=1$ equals the force acting between two charged particles separated by a distance $r=a$.

To derive the spatial form of the perturbed density profile we consider the perturbation series (\ref{sor}) truncated to second order:
\begin{eqnarray}
\langle\tilde{n}(\textbf{k}_0)\rangle=\hat{\chi}^{(1)}(\textbf{k}_0)\hat{\Phi}(\textbf{k}_0)+ \nonumber \\
\frac{1}{V}\sum_{\textbf{p}_{1}}\hat{\chi}^{(2)}(\textbf{p}_{1},-\textbf{k}_0-\textbf{p}_{1})\hat{\Phi}(\textbf{p}_{1})\hat{\Phi}(-\textbf{p}_1-\textbf{k}_{0}). 
\label{sor2}
\end{eqnarray}
Using
\begin{equation}
\hat{\Phi}(\textbf{k})=\frac{c f_{0}V}{2}\Big[\delta_{\textbf{k},-\textbf{k}_{1}}+\delta_{\textbf{k},\textbf{k}_{1}}\Big],
\label{Fourier}
\end{equation}
(\ref{sor2}) gives, after inverse Fourier transform:
\begin{eqnarray}
\langle\tilde{n}(\textbf{r})\rangle = ~~~~~~~~~~~~~~~~~~~\nonumber \\
\frac{cf_{0}}{2}\Big[\hat{\chi}^{(1)}(\textbf{k}_{1})e^{i \textbf{k}_{1}\textbf{r}}+\hat{\chi}^{(1)}(-\textbf{k}_{1})e^{- i \textbf{k}_1 \textbf{r}}\Big]+ \nonumber \\
\frac{c^{2}f_{0}^{2}}{4} \bigg\{ \Big[\hat{\chi}^{(2)}(-\textbf{k}_{1},\textbf{k}_{1})+\hat{\chi}^{(2)}(\textbf{k}_{1},-\textbf{k}_{1})\Big]+
\nonumber \\
\hat{\chi}^{(2)}(-\textbf{k}_{1},-\textbf{k}_{1})e^{i2\textbf{k}_{1}\textbf{r}}+ \hat{\chi}^{(2)}(\textbf{k}_{1},\textbf{k}_{1})e^{-i2\textbf{k}_{1}\textbf{r}}\bigg\}. \label{sor2v}
\end{eqnarray}
The QFDT requires that the response function remain invariant with respect to the permutation of all their arguments (${\bf k}_1,{\bf k}_2, {\bf k}_0$), i.e. they obey a triangle symmetry. Making use of this invariance and the invariance of the response functions with respect to spatial reflection, eq. (\ref{sor2v}) can be rewritten to the following form:
\begin{eqnarray}
\langle\tilde{n}(\textbf{r})\rangle=cf_{0}\hat{\chi}^{(1)}(\textbf{k}_{1})\cos(\textbf{k}_{1}\textbf{r})+\nonumber\\
\frac{c^{2}f_{0}^{2}}{2}\Bigl[\hat{\chi}^{(2)}(\textbf{k}_{1},-\textbf{k}_{1})+ \hat{\chi}^{(2)}(\textbf{k}_{1},\textbf{k}_{1})\cos(2\textbf{k}_{1}\textbf{r}) \Bigr].
\label{sor2vcos}
\end{eqnarray}

We note that, following a similar approach, and expecting that the higher order response functions obey a similar permutation symmetry, the higher order contributions to the perturbed density profile can be derived as well. Here we give these contributions (without their derivation), up to 4-th order:
\begin{eqnarray}
\langle\tilde{n}(\textbf{r})\rangle^{(1)}&=&cf_{0}\hat{\chi}^{(1)}(\textbf{k}_{1})\cos(\textbf{k}_{1}\textbf{r})\label{1rend}\\
\langle\tilde{n}(\textbf{r})\rangle^{(2)}&=&\frac{c^{2}f_{0}^{2}}{2}\Big[\hat{\chi}^{(2)}(\textbf{k}_{1},-\textbf{k}_{1})+ \nonumber \\ && \hat{\chi}^{(2)}(\textbf{k}_{1},\textbf{k}_{1})\cos(2\textbf{k}_{1}\textbf{r})\Big]\label{2rend}\\
\langle\tilde{n}(\textbf{r})\rangle^{(3)}&=&\frac{c^{3}f_{0}^{3}}{4}\Big[3\hat{\chi}^{(3)}(\textbf{k}_{1},\textbf{k}_{1},-\textbf{k}_{1})\cos(\textbf{k}_{1}\textbf{r})+ \nonumber \\ &&\hat{\chi}^{(3)}(\textbf{k}_{1},\textbf{k}_{1},\textbf{k}_{1})\cos(3\textbf{k}_{1}\textbf{r})\Big]\label{3rend}\\\nonumber
\langle\tilde{n}(\textbf{r})\rangle^{(4)}&=&\frac{c^{4}f_{0}^{4}}{8}\Big[3\hat{\chi}^{(4)}(\textbf{k}_{1},\textbf{k}_{1},-\textbf{k}_{1},-\textbf{k}_{1})+ \nonumber \\ && 4\hat{\chi}^{(4)}(\textbf{k}_{1},-\textbf{k}_{1},-\textbf{k}_{1},-\textbf{k}_{1})\cos(2\textbf{k}_{1}\textbf{r})+ \nonumber \\
&&\hat{\chi}^{(4)}(\textbf{k}_{1},\textbf{k}_{1},\textbf{k}_{1},\textbf{k}_{1})\cos(4\textbf{k}_{1}\textbf{r})\Big]\label{4rend}
\end{eqnarray}

The first order term (\ref{1rend}) is used for the calculation of the linear response function $\hat{\chi}^{(1)}$, while the second order term (\ref{2rend}) is used for the calculation of the quadratic response function $\hat{\chi}^{(2)}$ in the diagonal direction (i.e. having identical wave numbers as the arguments). The second order term contains the second harmonic with wave number vector ${\bf k}_0 = 2 {\bf k}_1$, as well as a ``DC contribution'' $\hat{\chi}^{(2)}(\textbf{k}_{1},-\textbf{k}_{1})$. For this latter term ${\bf k}_0= 0$, so it belongs to the domain of singular behavior (see our comment at the end of Sec. IIA). Also, as the conservation of particle number in the simulation excludes the appearance of any DC contribution, this term (and similar higher order counterparts) will be ignored.

\subsection{Biharmonic perturbation}

Here we investigate the effect of two independent perturbations with different wave numbers (but equal amplitudes) acting simultaneously on the system, i.e.
\begin{equation}
\hat{\Phi}(\textbf{r})=\frac{c f_0}{2} \big[ \cos(\textbf{k}_1\textbf{r}) +
\cos(\textbf{k}_2 \textbf{r}) \big].
\end{equation}
The perturbed density profile can be determined using the same approach as detailed above. The density profile up to second order is the following:

\begin{eqnarray}
\langle\tilde{n}(\textbf{r})\rangle= \frac{cf_{0}}{2}\Big[\hat{\chi}^{(1)}(\textbf{k}_1) \cos\big(\textbf{k}_1\textbf{r}\big)+
\hat{\chi}^{(1)}(\textbf{k}_2) \cos\big(\textbf{k}_2 \textbf{r}\big)\Bigr] +\nonumber \\
\frac{c^{2}f_{0}^{2}}{8}\Big[\hat{\chi}^{(2)}(\textbf{k}_1,-\textbf{k}_1)+
\hat{\chi}^{(2)}(\textbf{k}_2,-\textbf{k}_2)\Big]+
\nonumber \\
\frac{c^{2}f_{0}^{2}}{4}\Big[\hat{\chi}^{(2)}(\textbf{k}_1,\textbf{k}_2)\cos\big((\textbf{k}_1+\textbf{k}_2) \textbf{r}\big)+ \nonumber \\
\hat{\chi}^{(2)}(\textbf{k}_1,-\textbf{k}_2)\cos\big((\textbf{k}_1-\textbf{k}_2)\textbf{r}\big)\Big]+\nonumber \\
\frac{c^{2}f_{0}^{2}}{8}\Big[\hat{\chi}^{(2)}(\textbf{k}_1,\textbf{k}_1)\cos\big(2 \textbf{k}_1 \textbf{r}\big)+\nonumber \\
\hat{\chi}^{(2)}(\textbf{k}_2,\textbf{k}_2)\cos\big( 2 \textbf{k}_2 \textbf{r}\big)\Big]~~~~~~~~
\label{sor2vbi}
\end{eqnarray}

The first term in this equation is the sum of the linear responses to the two parts of the perturbing potential. Similarly to the single harmonic case, the next term would give a DC contribution, however this term is ignored based on the arguments given above. It is the third term, containing $\cos\big( (\textbf{k}_{1} \pm \textbf{k}_{2}) \textbf{r} \big)$, that can be used in the ``measurements'' of the quadratic response function $\hat{\chi}^{(2)}$ with arguments $\textbf{k}_{1} \neq \textbf{k}_{2}$. Finally, the last term would allow the determination of $\hat{\chi}^{(2)}$ in the diagonal direction, which can be done as well using the single harmonic perturbation, as described in the previous subsection.

\section{Simulations}

\subsection{Equilibrium and non-equilibrium simulations}

Equilibrium and non-equilibrium molecular dynamics (MD)  simulations have been used extensively in the calculations of static properties, transport coefficients, collective excitations, as well as instabilities in strongly coupled Yukawa liquids (see e.g.\cite{m1,m2,Murillo,m3,m4,m5,m6,m7,m8,m9,m10,m11,m12}).  Here we use a standard MD method to describe our 3-dimensional Yukawa liquid \cite{FS}: we simulate the motion of $N$=16,000 particles, within a cubic box with periodic boundary conditions, via the integration of their Newtonian equations of motion. The spatial decay of the Yukawa interaction makes it possible to introduce a cutoff distance in the calculation of the forces acting on the particles, beyond which the interaction of particle pairs can be neglected. For our conditions $r_{\rm cutoff} \approx 10.2 a$. Time integration is performed using the velocity-Verlet scheme. At the initialization of the simulation runs the positions of the particles are set randomly, while their initial velocities are sampled from a Maxwellian distribution corresponding to a specified system temperature. The simulations start with a thermalization phase, during which the particle velocities are rescaled in each time step, in order to reach the desired temperature. This procedure is stopped before the data collection takes place, where the stability of the simulation is confirmed by monitoring the temperature as a function of time.

We employ two types of simulations: with and without external perturbation. The equilibrium MD (EMD) simulations (on the unperturbed system) are used for the computation of the pair correlation function $h(r)$, as well as the static structure factor $S({\bf k}_1)$ and its quadratic counterpart $S({\bf k_1},{\bf k_2})$. These quantities are derived from the phase space trajectories of the particles according to equations (\ref{eq:linsk}) and (\ref{eq:quadsk}). In the non-equilibrium MD (NEMD) simulation runs the (harmonic or biharmonic) external potential is applied from the beginning of the simulation, both during the thermalization and measurement phases. In these simulations the primary target is to measure the spatial density distribution of the particles, $\langle n({\bf r})\rangle$. 

In all investigated cases the wave number dependence of the response functions is scanned in a sequence of simulation runs, in a way, that the amplitude of the perturbation ($f_0$) is varied to result for each $k$ in a $\cong$10\% modulation of the density profile via the linear response. The contribution of the quadratic response is typically one order of magnitude smaller, but the quadratic response functions can be still determined from these density profiles with an acceptable accuracy. The limitations of our method appear at low wave numbers, where the response of the system is weak, and, therefore, a high amplitude of the external potential energy needs to be used to induce an appreciable density modulation. For these conditions we have observed that $s>2$ order responses also contribute to the induced density profiles, and invalidate the assumptions used in our data analysis procedures. Our data analysis still gives $\hat{\chi} \cong 0$ values in this domain, as the linear response is weak, but the accuracy of the data here is inferior, compared to that in the domain of higher wave numbers, where an appreciable response appears. Analysis of the  responses of different orders as a function of the perturbation amplitude $f_0$ at selected values of the wave number confirmed the correctness of our data acquisition procedure in the $k a \gtrsim 2$ range.

\subsection{Harmonic perturbation}

\begin{figure}[htb]
\includegraphics[width=0.4\columnwidth]{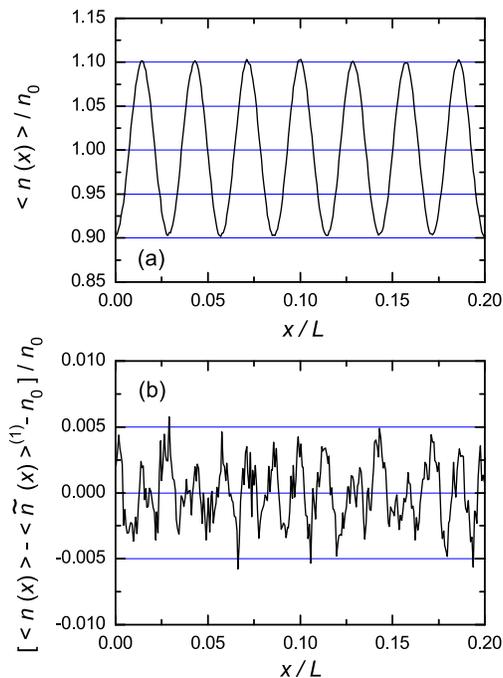}
\caption{\label{fig:pert1}
The normalized density distribution of the system in case of a harmonic external potential with ${\bf k}_1 a = (35,0,0) k_{min}a$. (a) Total density response, (b) the nonlinear part of the response. The plots show only a part of the simulation box.}
\end{figure}

The determination of the linear response function $\hat{\chi}^{(1)}({\bf k}_1)$ and the quadratic response function in the diagonal direction $\hat{\chi}^{(2)}({\bf k}_1,{\bf k}_1)$ is based on the expressions (\ref{1rend}) and (\ref{2rend}). In the MD simulations we set the wave number vector to point into the $x$ direction, i.e. ${\bf k}_1 a = (k_1,0,0)a$, and as a result we obtain the $\langle n(x)\rangle$ profile. In Fig.~\ref{fig:pert1} we illustrate this profile for ${\bf k}_1 a = (35,0,0) k_{min}a$, where $k_{min}a = 2 \pi a/L = 0.155$ is the minimum accessible wave number defined by the edge length $L$ of the simulation box. The amplitude of the external potential energy was set in a way to result a $\cong$10\% modulation of the density profile, as explained above.

The density response shown in Fig.~\ref{fig:pert1}(a) is very close to harmonic, but a slight asymmetry can be observed: the maximum positive deviation from uniform density is somewhat higher than the maximum negative deviation. Fig.~\ref{fig:pert1}(b) shows the nonlinear part, that is the sum of higher order responses, dominated by (and is assumed in our data acquisition procedures to originate exclusively from) the second order response. This profile is quite noisy, despite the fact that averaging of the density profiles proceeds during $200,000$ time steps in the simulations. Nonetheless, the amplitudes of the first and second harmonic content of the (total) $\langle n(x) \rangle$ distribution can be obtained with reasonable accuracy, as this procedure involves a spatial integration of the profile.

The linear and quadratic response functions (assuming that higher-order terms have negligible contributions due to the proper choice of $f_0$, that ensures an $\approx$ 10\% density perturbation via the linear term) are readily obtained as
\begin{eqnarray}
\hat{\chi}^{(1)}(k_1) &=& \frac{A_1 }{ c f_0},\\
\hat{\chi}^{(2)}(k_1,k_1) &=& \frac{2 A_2} {c^2 f^2_0},
\end{eqnarray}
where $A_1$ and $A_2$, respectively, are the amplitudes of the first and second harmonic contributions (mentioned above) to $\langle n(x) \rangle$. The full wave number dependence of $\hat{\chi}^{(1)}$ and $\hat{\chi}^{(2)}$ is scanned by carrying out a series of simulations with $k_1 = m k_{min}$, where $m=1,2,...,65$.

\begin{figure}[htb]
\includegraphics[width=0.4\columnwidth]{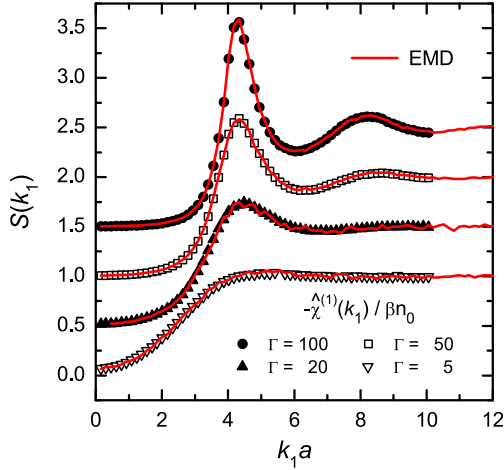}
\caption{\label{fig:sk}
(color online) Normalized linear response functions $-\hat{\chi}^{(1)}(k_1)/\beta n_0$ obtained from the NEMD simulations (symbols), in comparison with the static structure factors $S(k_1)$ resulting from an EMD simulations (lines) for $\Gamma$ values indicated, $\kappa=1$. The curves are vertically shifted for the clarity of the plot.}
\end{figure}

Figure~\ref{fig:sk} shows the normalized form of the linear response function, $-\hat{\chi}^{(1)}(k_1)/\beta n_0$, which, according to the linear FDT, has to equal the static structure function $S(k_1)$. Indeed, we find an excellent agreement between the two sets of data, obtained from the EMD, on the one hand, and from the NEMD simulations, on the other, for all the coupling parameter values covered. The calculation of the static structure function $S(k_1)$ in an alternative manner, via the Fourier transform of the equilibrium pair correlation function, $h(r)$, has yielded identical results.

\begin{figure}[t!]
\includegraphics[width=0.4\columnwidth]{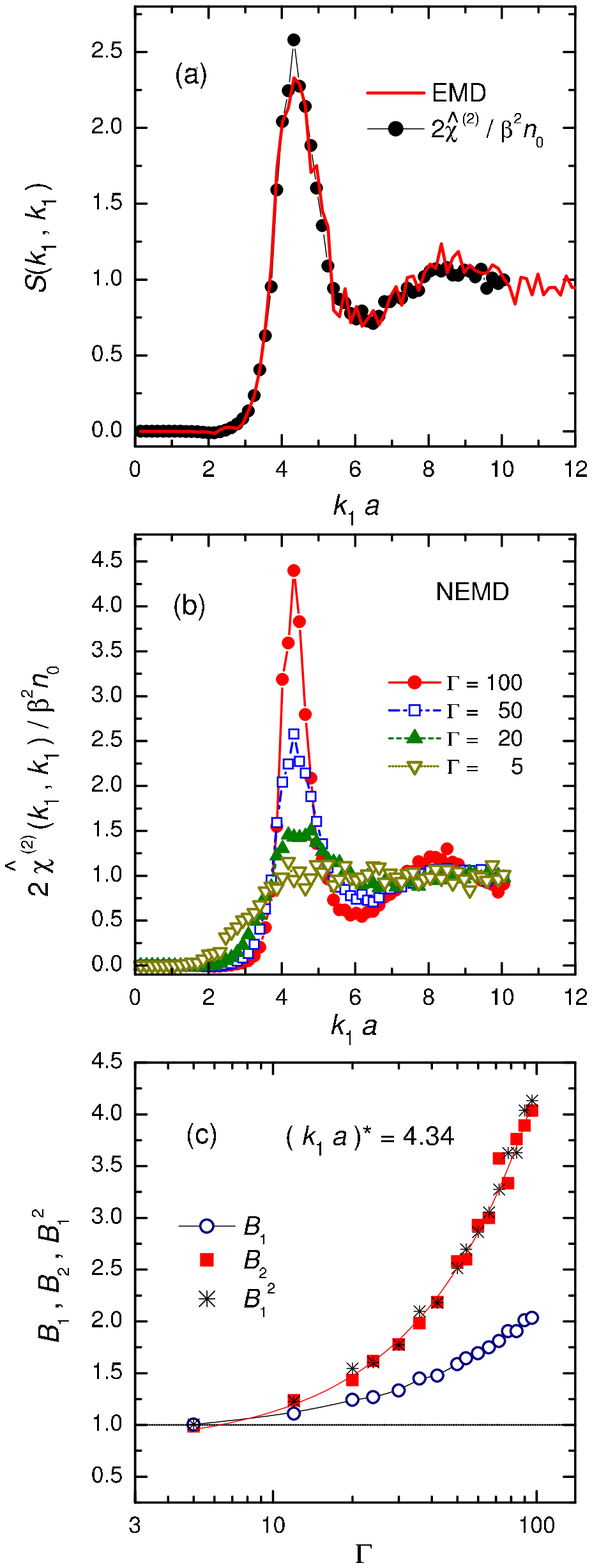}
\caption{\label{fig:skk}
(color online) (a) Normalized quadratic response function $2 \hat{\chi}^{(2)}(k_1,k_1)/\beta^2 n_0$ obtained from the NEMD simulations (symbols), in comparison with the static structure factor $S(k_1,k_1)$ resulting from an EMD simulation (line), for $\Gamma=50$, $\kappa=1$. (b) Dependence of $2 \hat{\chi}^{(2)}(k_1,k_1)/\beta^2 n_0$ on the coupling parameter $\Gamma$, at $\kappa=1$. (c) Amplitudes $B_2$ of $2 \hat{\chi}^{(2)}(k_1,k_1)/\beta^2 n_0$ and $B_1$ of $S(k_1)$ [equivalent to $-\hat{\chi}^{(1)}(k_1)/\beta n_0$], as well as its square, $B_1^2$, as a function of $\Gamma$, at $(k_1 a)^\ast = 28 k_{min}a = 4.34$ (position of the first peak of the functions).}
\end{figure}

Figure~\ref{fig:skk}(a) shows, for $\Gamma=50$ and $\kappa =1$ the normalized form of the quadratic response function, $2 \hat{\chi}^{(2)}/\beta^2 n_0$, which, according to the quadratic FDT has to equal the static structure function $S(k_1,k_1)$. A convincing agreement between the two sets of data has been found here as well, although the quadratic data sets are more noisy compared to the linear case. This results from the facts that (i) the higher order perturbation of the density, $\langle\tilde{n}(\textbf{r})\rangle^{(2)}$, is about an order of magnitude smaller than the linear term, and that (ii) in the EMD the collection of the data for $S(k_1,k_1)$ is more time consuming compared to the case of $S(k_1)$. We note that about 3 years of CPU time was devoted (using $\sim$50 CPU-s to the direct generation of $S(k_1,k_1)$ (via the EMD)). The generation of each of the 65 data points with the NEMD took 5 days of CPU time (total CPU time $\sim$ 1 year). Comparison of these run times indicates that the NEMD is more efficient to generate the quadratic response function, compared to the calculation that proceeds via the equilibrium simulations for the static structure function. Therefore, we have only used the NEMD method to study the dependence of $\chi^{(2)}(k_1,k_1)$ on $\Gamma$, as shown in Fig.~\ref{fig:skk}(b). The amplitudes $B_1$ and $B_2$ of the linear and quadratic response functions, respectively, have been calculated for additional $\Gamma$ values at the peak position, $(k_1 a)^\ast \cong$ 4.34, and are presented in Fig.~\ref{fig:skk}(c). (The $B_1$ values originate from EMD calculations, while data for $B_2$ have been obtained in NEMD calculations.) Both amplitudes increase with increasing $\Gamma$, as expected, due to the more prominent structure at higher coupling values. Additionally, we find an unexpected agreement $B_1^2 \cong B_2$, which suggests that the quadratic structure function can be factorized in terms of linear structure functions. This property will be examined in more details later.

\subsection{Biharmonic perturbation}

Using a biharmonic external potential energy $\hat{\Phi}(\textbf{r})=\hat{\Phi}_{0}[\cos(\textbf{k}_{1}\textbf{r})+\cos(\textbf{k}_{2}\textbf{r})]$ allows us to determine the quadratic response function in non-diagonal directions. To determine the response function we measure the amplitude of the following harmonic term of (\ref{sor2vbi}):
\begin{equation}
\frac{c^{2}f_{0}^{2}}{4} \hat{\chi}^{(2)}(\textbf{k}_1,\textbf{k}_2)\cos\big((\textbf{k}_1+\textbf{k}_2) \textbf{r}\big). 
\end{equation}
In the following we shall discuss two cases.

The first is a special case when the wave number arguments of $\hat{\chi}^{(2)}$ are parallel, and ${\bf k}_2=2{\bf k}_1$. To generate a proper response of the system we apply an external potential with wave vectors directed into the $x$ direction, and use scalar quantities correspondingly, i.e. $k_2=2k_1$, with $k_1 a = m~k_{min} a$. 

\begin{figure}[t!]
\includegraphics[width=0.4\columnwidth]{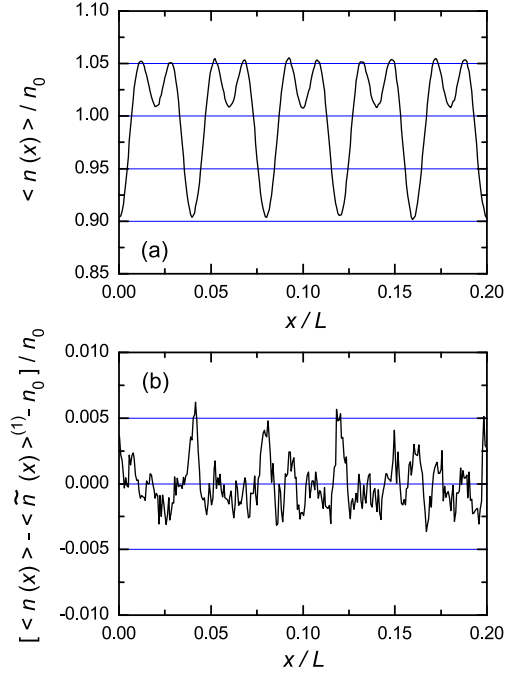}
\caption{\label{fig:pert2}
(color online) (a) Perturbed density distribution of the system in case of a biharmonic external potential with ${k}_1a =25 k_{min}a $ and $k_2 a = 2 k_1 a$. (b) The nonlinear part of the perturbed density. The plots show only a part of the simulation box.}
\end{figure}

The emerging perturbed density distribution for $k_1 a =25 k_{min} a$ is illustrated in Fig.~\ref{fig:pert2}. The perturbation of the density distribution is dominated by the linear contributions, according to the first term of (\ref{sor2vbi}). It is the small deviation (similar in relative magnitude to that seen in Fig.~\ref{fig:pert1}) of $\langle \tilde{n}(x) \rangle$ from this term that is due to the second order response [see Fig.~\ref{fig:pert2}(b).]

\begin{figure}[h!]
\includegraphics[width=0.4\columnwidth]{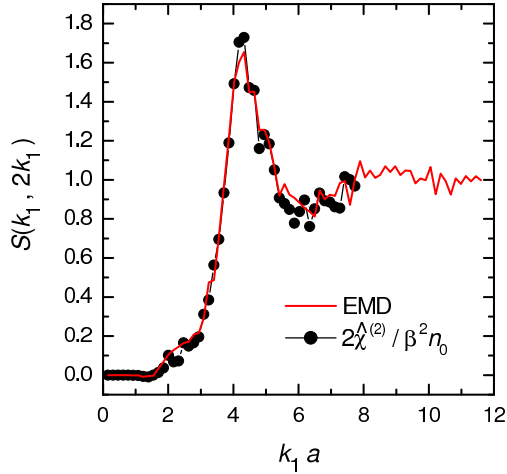}
\caption{\label{fig:sk2k}
(color online) Normalized quadratic response function $2 \hat{\chi}^{(2)}(k_1,2k_1)/\beta^2 n_0$ obtained from the NEMD simulations (symbols), in comparison with the static structure factor $S(k_1,2k_1)$ resulting from an EMD simulation. $\Gamma=50$, $\kappa=1$. }
\end{figure}

The resulting normalized quadratic response function $2 \hat{\chi}^{(2)}(k_1,2k_1)/\beta^2 n_0$ obtained from the NEMD simulations, in comparison with the static structure factor $S(k_1,2k_1)$ resulting from an equilibrium simulation is plotted in Fig.~\ref{fig:sk2k}, for $\Gamma=50$ and $\kappa=1$. A good agreement between the two data sets is obtained in this case as well, as in the diagonal case analyzed earlier. Note, however, the remarkable feature that $2 \hat{\chi}^{(2)}(k_1,2k_1)/\beta^2 n_0$ is quite similar to the $2 \hat{\chi}^{(2)}(k_1,k_1)/\beta^2 n_0$  in Fig.~\ref{fig:skk}(a). This feature further suggests that the quadratic structure function can be factorized in terms of linear structure functions.

\begin{figure}[htb]
\includegraphics[width=0.4\columnwidth]{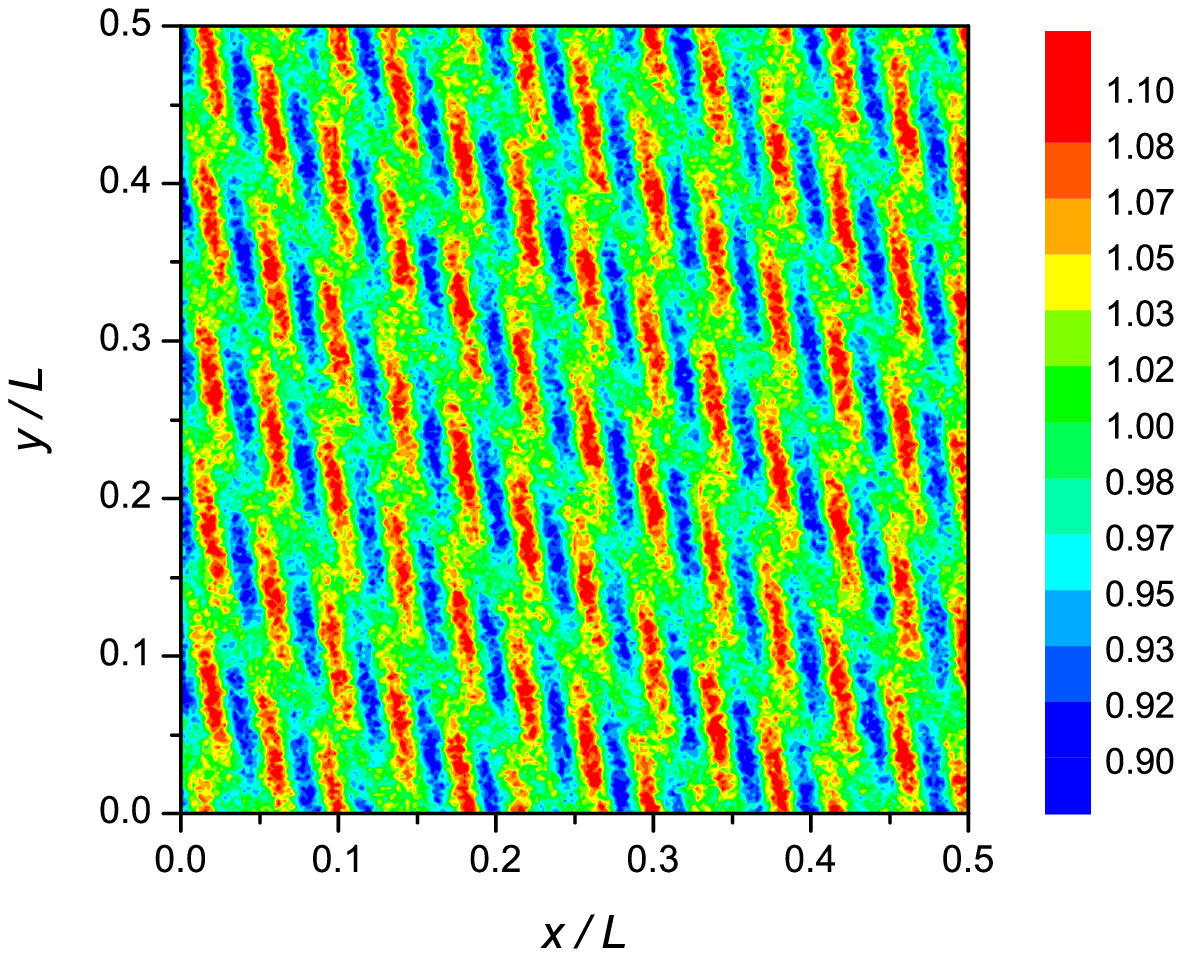}
\caption{\label{fig:pert3}
(color online) The perturbed density distribution of the system in case of a biharmonic external potential, at ${\bf k}_1 a=(35,10,0)k_{min}a$ and ${\bf k}_2a = (25,0,0)k_{min}a$. $\Gamma=50$, $\kappa=1$.}
\end{figure}

\begin{figure}[htb]
\includegraphics[width=0.4\columnwidth]{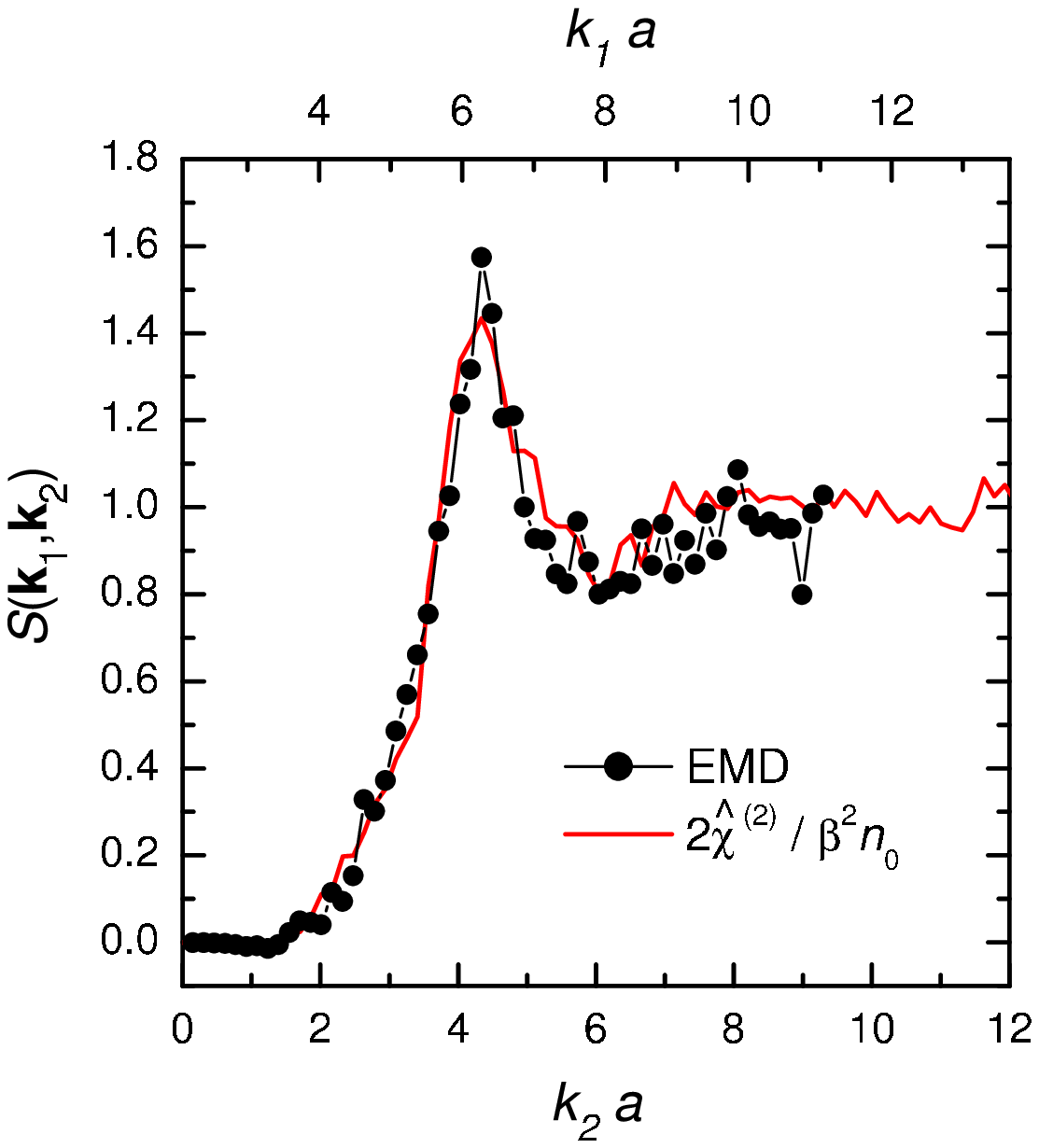}
\caption{\label{fig:sk1k2}
(color online) Normalized quadratic response function $2 \hat{\chi}^{(2)}(\textbf{k}_{1},\textbf{k}_{2})/\beta^2 n_0$ obtained from the NEMD simulations, in comparison with the static structure factor $S(\textbf{k}_{1},\textbf{k}_{2})$ resulting from an EMD simulation. $\textbf{k}_{1}a=(m+5,10,0)k_{min}a$ and $\textbf{k}_{2}a=(m-5,0,0)k_{min}a$,  $\Gamma=50$, $\kappa=1$. }
\end{figure}

In the second, more general case we take $\textbf{k}_{1}a=(m+5,10,0)k_{min}a$ and $\textbf{k}_{2}a=(m-5,0,0)k_{min}a$, with $m$ being an integer number. The angle of the two wave number vectors, as a function of $m$ ($m \neq \pm 5 $) varies as $\alpha = \arccos (1/\sqrt{1+100/(m+5)^2})$.  Figure~\ref{fig:pert3} displays, as an example, the emerging perturbed density distribution for ${\bf k}_1a =(35,10,0)k_{min}a$ and ${\bf k}_2a =(25,0,0)k_{min}a$. The resulting normalized quadratic response function $2 \hat{\chi}^{(2)}(\textbf{k}_{1},\textbf{k}_{2})/\beta^2 n_0$ obtained from the NEMD simulations, in comparison with the static structure factor $S(\textbf{k}_{1},\textbf{k}_{2})$ resulting from an EMD simulation is plotted in Figure~\ref{fig:sk1k2}. A good agreement is again found between the two sets of data, verifying the quadratic FDT for a more general case.

Errors of the calculated response functions originate from two sources: (i) from omitting of the contributions of higher-order response functions to the perturbed density profiles used for the determination of the linear and quadratic response, and (ii) from the statistical noise of the simulations. The first of these sources is estimated to be at the 1\% level, ensured by using a low degree of perturbation in the $k a \gtrsim 2$ domain. The statistical errors of the simulation are estimated to be at the 1\% level in the case of the first order structure functions and responses, and to be at the 10\% level in the case of the second order structure functions and responses. Improving the latter is a technical issue, using a bigger system size and longer simulations the statistical noise can be suppressed. 

\begin{figure}[htb]
\includegraphics[width=0.4\columnwidth]{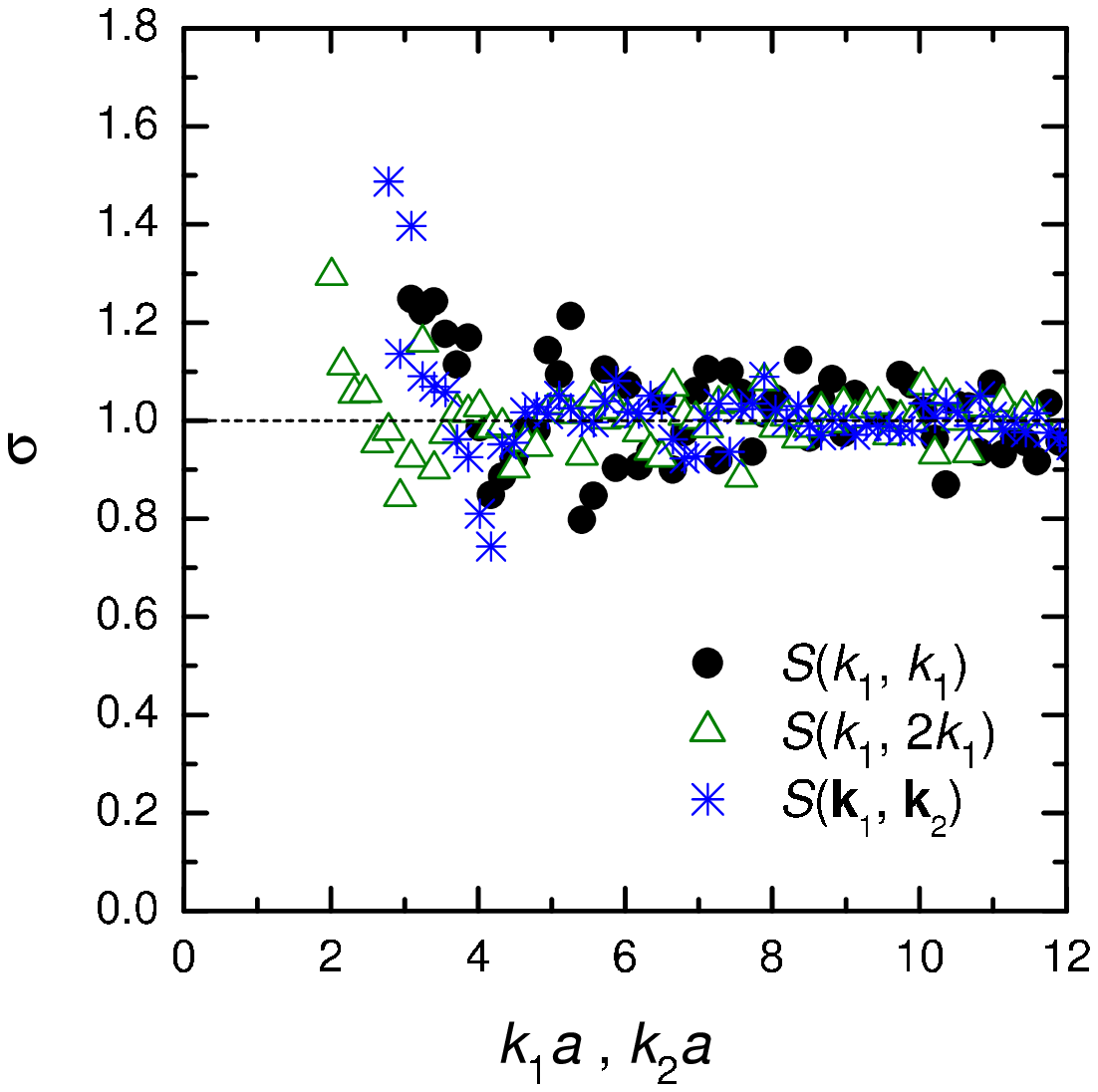}
\caption{\label{fig:sigma}
(color online) Factorization ratio $\sigma$ as a function of wave number, for $\Gamma=50$, $\kappa=1$. In the case of $S(k_1,k_1)$ and $S(k_1,2k_1)$ the data are plotted as a function of $k_1a$, while for the more general case $S({\bf k}_1,{\bf k}_2)$ the data are plotted as a function of $k_2 a$.}
\end{figure}

\subsection{Factorization and correlation functions}

We have noted before that the structures found in the cases studied suggest that $S({\bf k}_1,{\bf k}_2;{\bf k}_0)$ can be factorized in terms of linear structure functions:
\begin{equation}
\label{eq:fact}
S({\bf k}_1,{\bf k}_2; {\bf k}_0) \cong S({\bf k}_1) S({\bf k}_2) S({\bf k}_0),
\end{equation}
with ${\bf k}_0+{\bf k}_1+{\bf k}_2 = 0$. To see this, we calculate the ``factorization ratio''
\begin{equation}
\sigma({\bf k}_1,{\bf k}_2) = \frac{S({\bf k}_1,{\bf k}_2; {\bf k}_0 )} {S({\bf k}_1) S({\bf k}_2) S({\bf k}_0)}
\end{equation}
for the cases examined above: (i) $S(k_1,k_1)$, (ii) $S(k_1,2k_1)$, and (iii) $\textbf{k}_{1}a=(m+5,10,0)k_{min}a$ and $\textbf{k}_{2}a=(m-5,0,0)k_{min}a$. [In cases (i) and (ii) the wave number points in the $x$ direction, as earlier.] The data obtained for $\sigma$ in EMD simulations are plotted in Fig.~\ref{fig:sigma}. Despite the relatively large scatter of the data it is obvious that the factorization ratio is nearly 1.0. These data unambiguously show that the quadratic structure function can be factorized in terms of linear structure functions, at least for the given coupling-screening parameter pair. However, a few caveats are in order. First, the increasing deviation from $\sigma \cong 1$ towards low wave numbers may originate from the lack of accurate data at small $k$ values, aggravated by the division by the linear $S(k)$-s, which are close to zero at $k \rightarrow 0$ and, therefore, are rather unreliable. Nevertheless, there seems to be a trend towards the breakdown of the factorization for small wave numbers. Second, there are theoretical constraints that limit the factorizability of $S({\bf k}_1,{\bf k}_2;{\bf k}_0)$: foremost amongst these is the quadratic compressibility sum rule (QCSR) established in \cite{Datta}. This sum rule that applies in the ${\bf k}_1 \rightarrow 0$, ${\bf k}_2 \rightarrow 0$, ${\bf k}_0 \rightarrow 0$ limit, can be compared with the behavior ensuing from the factorized expression, governed by the conventional compressibility sum rules obeyed by the linear $S({\bf k}_1)$, etc. As shown in the Appendix in more detail, what ensues from these considerations is that factorization in the ${\bf k} \rightarrow 0$, limit is possible only as long as the equation of state (EOS) contains $O(n)$ and $O(n^2)$ terms only. The former corresponds to the case of the perfect gas, the second encompasses the Hartree EOS, which is the leading term in the EOS of a Yukawa system.  Correlational contribution to the EOS, that for high $\Gamma$ values scales as $\Gamma n$ \cite{Hartmann}, i.e. $O(n^{4/3})$ would violate the  QCSR condition. 

Finally, even though the factorization formula (\ref{eq:fact}) is not entirely new (it can be readily shown from \cite{AnnPhys} to be exact in the Random Phase Approximation, RPA, limit), here we are dealing with a strongly correlated system whose behavior is not described by the RPA. 

The  factorization property in conjunction with (14) leads to a closed (albeit approximate) expression for the hitherto unknown $h({\bf r}_{12},{\bf r}_{23})$ irreducible three-particle correlation function for a strongly coupled Yukawa system. From
	 
\begin{align}
h(k_1,k_2) =  h(k_1) h(k_2) + h(k_2) h(k_0) & \nonumber \\
+ h(k_0) h(k_1) + n \big[ h(k_1) h(k_2) h(k_0) \big]&
\label{eq:hk1k2}
\end{align} 
we find 
\begin{eqnarray}
h({\bf r}_{12},{\bf r}_{23}) =  ~~~~~~~~~~~~~&\\ 
n \Biggl[ \int {\rm d}^3 {\bf r}_4 h({\bf r}_{14})h({\bf r}_{42})+ \int {\rm d}^3 {\bf r}_4 h({\bf r}_{24})h({\bf r}_{43})+ & \nonumber \\
\int {\rm d}^3 {\bf r}_4 h({\bf r}_{34})h({\bf r}_{41})+  \int {\rm d}^3 {\bf r}_4 h({\bf r}_{14}) h({\bf r}_{24}) h({\bf r}_{34}) \Biggr] . &\nonumber
\end{eqnarray} 

To the best of our knowledge, the above cluster representation is the first reliable information obtained for the structure of $h({\bf r}_{12},{\bf r}_{23})$ in a strongly coupled environment. Note the absence of simple Kirkwood-like $h({\bf r}_{12}) h({\bf r}_{23})$, etc., or $h({\bf r}_{12}) h({\bf r}_{23}) h({\bf r}_{31})$ products. The expected breakdown of the factorization approximation in the small $k$ domain would most likely translate into an error in the cluster representation of $h({\bf r}_{12},{\bf r}_{23})$ in the $r \rightarrow \infty$ limit, where triplet correlations are already weak. The dependence on $\Gamma$ and $\kappa$ has not been investigated, a study that covers a wide range of these parameters is planned to be carried out.

\section{Summary}

In this paper we have investigated the linear and nonlinear density responses of 3-dimensional strongly coupled Yukawa liquids to external potential perturbations. The response functions were determined in non-equilibrium molecular dynamics (NEMD) simulations. Applying a single harmonic perturbation allowed to measure the linear response function $\hat{\chi}^{(1)}(k)$ and the quadratic response function in the diagonal direction in wave number space, $\hat{\chi}^{(2)}(k,k)$. Using a biharmonic perturbation allowed the determination of the quadratic response function with arbitrary arguments, $ \hat{\chi}^{(2)}(\textbf{k}_{1},\textbf{k}_{2})$. 

Parallel to the NEMD simulations we have also carried out equilibrium (EMD) simulations to determine the static structure functions $S(k_1)$ and $S(\textbf{k}_{1},\textbf{k}_{2})$, linked with the response functions via the  linear and quadratic Fluctuation-Dissipation Theorems (FDT). The agreement of the results in the linear case verified our NEMD simulation method, while the agreement of the results in the quadratic case confirmed the quadratic FDT.  

At $\Gamma=50$ and $\kappa=1$ pair of parameters where we have performed a detailed study of the $k$-dependence of $S(k_1, k_2)$. It was found that the quadratic structure functions can be factorized in terms of linear structure functions. As a result, we have obtained a closed expression for the irreducible three-particle correlation function $h(r_{12},r_{23})$ in terms of cluster integrals of the two-particle correlation functions.

The NEMD approach has proven to be computationally more efficient in generating response functions of strongly coupled plasmas, compared to the EMD approach that proceeds via the generation of equilibrium static structure functions. Our method is also applicable to the calculation of higher ($> 2$) order response functions, however, such computations are foreseen to be rather demanding due to the decaying amplitudes of the higher order contributions to the perturbed density profiles. 

\begin{acknowledgments}
This work has been supported by the Grant OTKA K-105476 and NSF Grants  0813153, 0715224, 1105005.
We thank Dr P. Hartmann for useful discussions.
\end{acknowledgments}

\section{Appendix}

The compressibility sum rules for the linear screened and external density response functions state that
\begin{equation}
\chi(k \rightarrow 0) = - \frac{n_0}{K},
\tag{A1}
\end{equation}
\begin{equation}
\hat{\chi}(k \rightarrow 0) = - \frac{n_0}{K \varepsilon(k\rightarrow 0)},
\tag{A2}
\end{equation}	
where $K=(\partial P / \partial n_0)_T$ is the inverse isothermal compressibility.  The compressibility rule \cite{Datta} for the quadratic screened density response function is
\begin{equation}
\chi(k_1 \rightarrow 0, k_2 \rightarrow 0) = \frac{n_0}{2K^2} \Biggl[ 1-\frac{n_0}{K} \frac{\partial K}{\partial n_0} \Biggr].
\tag{A3}
\end{equation}	
Conversion to the external response function $\hat{\chi}(k_1 \rightarrow 0, k_2 \rightarrow 0)$ and trading that for $S(k_1 \rightarrow 0, k_2 \rightarrow 0; k_0 \rightarrow 0)$, via the QFDT gives
\begin{align}
&S(k_1 \rightarrow 0, k_2 \rightarrow 0; k_0 \rightarrow 0) = \frac{1}{\beta^2K^2} \Biggl[ 1-\frac{n_0}{K} \frac{\partial K}{\partial n_0} \Biggr] \times \nonumber \\
&\frac{1}{\varepsilon(k_0 \rightarrow 0) \varepsilon(k_1 \rightarrow 0) \varepsilon(k_2 \rightarrow 0)}. \tag{A4}
\end{align}	
Equating the left-hand-side of Eq. (A3) to the triple cluster formula 
\begin{align}
S(k_0 \rightarrow 0) S(k_1 \rightarrow 0) S(k_2 \rightarrow 0) = \nonumber \\
\frac{1}{\beta^3 K^3 \varepsilon(k_0 \rightarrow 0) \varepsilon(k_1 \rightarrow 0) \varepsilon(k_2 \rightarrow 0)}
\tag{A5}
\end{align}
[derived from (A2) and the linear FDT] then yields the consistency requirement:
\begin{equation}
\beta \frac{\partial K}{\partial n_0} = \frac{\beta}{n_0} K - \frac{1}{n_0}.
\tag{A6}
\end{equation}
This condition is satisfied insofar as the equation of state can be approximated as
\begin{equation}
P=C_0 n + C_1 n^2,
\tag{A7}
\end{equation}
which is certainly correct  within the Hartree approximation.  For any other structure, the factorization must be regarded as a (probably reasonable) approximation.

\end{document}